\documentclass[footinbib,aps,pra,reprint,superscriptaddress]{revtex4-1}

\usepackage{amsmath,amssymb,braket,upgreek,bm,verbatim}
\usepackage{graphicx,tabularx,xcolor} 
\definecolor{darkblue}{rgb}{0,0,0.5}
\usepackage{hyperref}
\hypersetup{
colorlinks=true,
linkcolor=black,
filecolor=blue,
citecolor=darkblue,  
urlcolor=black,
}

\usepackage[absolute]{textpos}
\usepackage{xcolor}
\usepackage{multirow}

\newcommand{\imag}{\mathrm{i}}

\usepackage{amsmath,amssymb,bbm,bm}
\usepackage{mathtools}
\usepackage{braket}
\usepackage{amsmath}
\usepackage{relsize}
\usepackage{comment}
\usepackage{csquotes}

\usepackage{accents}

\makeatletter
\ams@newcommand{\iiiiint}{\DOTSI\protect\MultiIntegral{5}}
\renewcommand{\MultiIntegral}[1]{%
  \edef\ints@c{\noexpand\intop
    \ifnum#1=\z@\noexpand\intdots@\else\noexpand\intkern@\fi
    \ifnum#1>\tw@\noexpand\intop\noexpand\intkern@\fi
    \ifnum#1>\thr@@\noexpand\intop\noexpand\intkern@\fi
    \ifnum#1>4 \noexpand\intop\noexpand\intkern@\fi 
    \noexpand\intop
    \noexpand\ilimits@
  }%
  \futurelet\@let@token\ints@a
}
\makeatother

\usepackage{stackengine}
\newcommand{\op}[1]{\mathop{}\!\ensurestackMath{\stackon[-.95ex]{%
  \mathbf{#1}}{\smash{\mathbf{\hat{}}}}}}
\newcommand{\opdag}[1]{\mathop{}\!\op{#1}^{\dag}}

\newcommand{\appropto}{\mathrel{\vcenter{
  \offinterlineskip\halign{\hfil$##$\cr
    \propto\cr\noalign{\kern2pt}\sim\cr\noalign{\kern-2pt}}}}}

\begin{document}
\title{Time-resolved Hanbury Brown--Twiss interferometry of on-chip biphoton frequency combs using Vernier phase modulation}

\author{Karthik V. Myilswamy}
\thanks{These authors contributed equally to this work.}
\author{Suparna Seshadri}
\thanks{These authors contributed equally to this work.}
\affiliation{School of Electrical and Computer Engineering and Purdue Quantum Science and Engineering Institute, Purdue University, West Lafayette, Indiana 47907, USA}
\author{Hsuan-Hao Lu}
\affiliation{Quantum Information Science Section, Oak Ridge National Laboratory, Oak Ridge, Tennessee 37831, USA}
\author{Mohammed S. Alshaykh}
\affiliation{School of Electrical and Computer Engineering and Purdue Quantum Science and Engineering Institute, Purdue University, West Lafayette, Indiana 47907, USA}
\affiliation{Electrical Engineering Department, King Saud University, Riyadh 11421, Saudi Arabia}
\author{Junqiu Liu}
\author{Tobias J. Kippenberg}
\affiliation{Institute of Physics, Swiss Federal Institute of Technology Lausanne (EPFL), 1015 Lausanne, Switzerland}
\author{Andrew M. Weiner}
\affiliation{School of Electrical and Computer Engineering and Purdue Quantum Science and Engineering Institute, Purdue University, West Lafayette, Indiana 47907, USA}
\author{Joseph M. Lukens}
\email{lukensjm@ornl.gov}
\affiliation{Quantum Information Science Section, Oak Ridge National Laboratory, Oak Ridge, Tennessee 37831, USA}
\affiliation{Knowledge Enterprise, Arizona State University, Tempe, Arizona 85287, USA}

\begin{textblock}{13.3}(1.4,15)
\noindent\fontsize{7}{7}\selectfont \textcolor{black!30}{This manuscript has been co-authored by UT-Battelle, LLC, under contract DE-AC05-00OR22725 with the US Department of Energy (DOE). The US government retains and the publisher, by accepting the article for publication, acknowledges that the US government retains a nonexclusive, paid-up, irrevocable, worldwide license to publish or reproduce the published form of this manuscript, or allow others to do so, for US government purposes. DOE will provide public access to these results of federally sponsored research in accordance with the DOE Public Access Plan (http://energy.gov/downloads/doe-public-access-plan).}
\end{textblock}

\begin{abstract}
Biphoton frequency combs (BFCs) are promising quantum sources for large-scale and high-dimensional quantum information and networking systems. In this context, the spectral purity of individual frequency bins 
will be critical for realizing quantum networking protocols like teleportation and entanglement swapping. Measurement of the temporal auto-correlation function of the unheralded signal or idler photons comprising the BFC is a key tool for characterizing their spectral purity and in turn verifying the utility of the biphoton state for networking protocols. 
Yet the experimentally obtainable precision for measuring BFC correlation functions is often severely limited by detector jitter. The fine temporal features in the correlation function---not only of practical value in quantum information, but also of fundamental interest in the study of quantum optics---are lost as a result and have remained unexplored.  We propose a scheme to circumvent this challenge through electro-optic phase modulation, experimentally demonstrating time-resolved Hanbury Brown--Twiss characterization of BFCs  generated from an integrated $40.5$~GHz Si$_3$N$_4$ microring, up to a $3\times3$-dimensional two-qutrit Hilbert space. 
Through slight detuning of the electro-optic drive frequency from the comb's free spectral range, our approach leverages Vernier principles to magnify temporal features which would otherwise be averaged out by detector jitter. We demonstrate our approach under both continuous-wave and pulsed pumping regimes, finding excellent agreement with theory. Importantly, our method reveals not only the collective statistics of the contributing frequency bins but also their temporal shapes---features lost in standard fully integrated auto-correlation measurements.

\end{abstract}
\maketitle

\section{Introduction}

Advances in photonic integrated circuits have enabled the miniaturization of 
biphoton frequency combs (BFCs)~\cite{kues2017chip,imany201850,kues2019quantum,lu2022bayesian} and provide a route toward scalable production of quantum circuits. 
Frequency-bin encoding is particularly attractive as integrated BFCs can produce a large number of spectrally entangled bins 
that can be coherently controlled with off-the-shelf telecommunication components~\cite{lukens2017frequency,lu2019quantum}. The time-resolved study of statistical properties of such BFCs is therefore not only of fundamental interest to quantum optics, but also of key significance to quantum information: for example, coherence across bins is critical for high-dimensional entanglement~\cite{kues2017chip,imany201850,lu2022bayesian}, spectral purity within a bin underpins successful interference between independent biphotons~\cite{eckstein2011highly, kaneda2016heralded, khodadad2021spectral}, and spectro-temporal mode matching is required for a photon to interface with cavities and stationary qubits~\cite{gorshkov2008photon,Liu2014,guo2019high,myilswamy2022temporal}.
The measurement of correlation functions is widely employed for studying such coherence properties of various quantum light sources~\cite{Glauber1963-1, mandel1995optical, ou2017quantum}.

We focus on the second-order auto-correlation of the unheralded signal (or idler) photons generated via spontaneous four-wave mixing (SFWM) in integrated microring resonators (MRRs). Hanbury Brown--Twiss (HBT) interferometry~\cite{brown1954lxxiv, brown1956correlation} has been a standard tool for these measurements, both for SFWM~\cite{kues2017chip,liu2020high,vaidya2020broadband,samara2021entanglement} and spontaneous parametric downconversion (SPDC)~\cite{tapster1998photon,mauerer2009colors,blauensteiner2009photon,luo2015direct,luo2017temporal}. Yet detector jitter frequently limits the temporal precision of such measurements. While slow detectors that integrate over entire pulses still provide useful information---such as the number of Schmidt modes in a pulse-pumped entangled photon source~\cite{christ2011probing}---the more general case of fully time-resolved measurements remains important for obtaining a comprehensive picture of the quantum state. 

For integrated BFCs, the typical few-hundred megahertz linewidths of microring resonances has allowed for direct measurement of second-order auto-correlation functions of \emph{individual} comb lines~\cite{guo2017parametric,samara2021entanglement} and cross-correlation functions of individual signal-idler pairs~\cite{guo2017parametric,jaramillo2017persistent}. But free spectral ranges (FSRs) on the order of tens of gigahertz and beyond have prevented full temporal resolution for \emph{multiple} comb lines together. In this work, we introduce a practical technique for circumventing detector resolution through Vernier-inspired electro-optic phase modulation. By driving modulators at a frequency slightly detuned from the comb FSR and filtering out an appropriate subspace of nearly overlapping sidebands, multibin interference can be observed directly with typical ($\sim$100-ps jitter) single-photon detectors. 

We demonstrate our technique on both continuous-wave (CW) and pulsed BFCs from a 40.5~GHz FSR Si$_3$N$_4$ microring, 
rescaling correlations originally periodic at 24.7~ps (inverse FSR) to 1~ns (inverse detuning) so that they can be comfortably resolved with commercial superconducting nanowire single-photon detectors (SNSPDs). 
Our approach improves effective temporal resolution and enables direct experimental characterization of the fine temporal features in BFC correlation functions, offering insights into the nature of the quantum sources under test otherwise buried in the averaged-out histograms of slow detectors.


\section{Scheme and Experimental Setup}\label{Experimental Scheme}
Our inspiration for resolution-enhanced photonic correlation measurements builds on a large body of work in single-photon manipulation with electro-optic modulation~\cite{Kolchin2008, Sensarn2009a, Belthangady2010, Olislager2010}. Modulators with tens of gigahertz---even $>$100~GHz~\cite{Wang2018b, zhu2021integrated}---of bandwidth are readily available, implying the ability to interact with and probe temporal features of single photons on the order of 1--100~ps. Thus whenever one is limited by the $\gtrsim$100~ps jitters of typical single-photon detectors, electro-optic modulators can both create and uncover temporal quantum features otherwise concealed from direct detection. Along these lines, modulators have been successfully applied to shaping photons with time lensing~\cite{karpinski2017bandwidth, Wright2017, mittal2017temporal}, coincidence resolution enhancement~\cite{Harris2008, Belthangady2009, Lukens2015}, and nonlocal delay sensing~\cite{seshadri2022nonlocal}. For integrated BFCs in particular, electro-optic modulation of the output photons at a frequency equal to the FSR (or a subharmonic thereof) causes the initial bins to overlap in frequency and interfere, which has facilitated full quantum state reconstruction of on-chip frequency-bin-entangled quantum states even with slow detectors~\cite{kues2017chip,imany201850,lu2022bayesian}.

We instead consider a modulation frequency intentionally detuned from the FSR such that adjacent bins no longer overlap. Conceptually similar to Vernier spectroscopy---where the interference of two frequency grids with different FSRs is used to enhance rapid acquisition of high-resolution spectra over broad bandwidths
~\cite{Gohle2007, coddington2016dual, wang2020vernier, Gomes2021}---our approach generates independent combs from each input frequency bin, creating clusters around each original bin with a sub-comb bin spacing equal to the detuning amount. Spectral filtering of one such cluster outputs a rescaled comb at a smaller effective FSR, where each contributing line can be uniquely identified with a single parent bin in the original BFC, leaving a temporally magnified picture into the BFC whose characteristics are now observable by slower detectors.

\begin{figure*}[!htb]
  \centering
  \includegraphics[trim=0 200 0 0,clip,width=\textwidth]{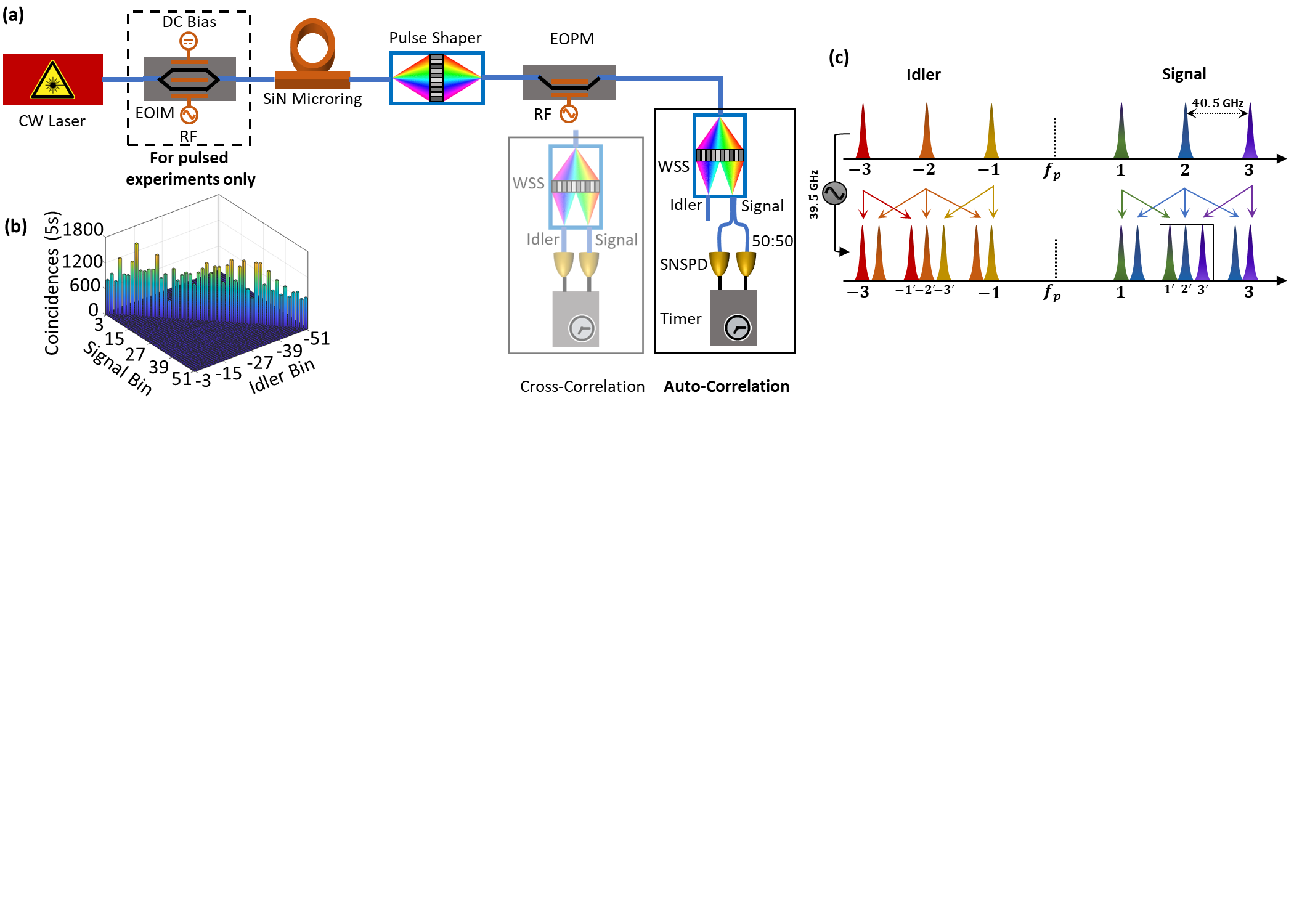}
\caption{
(a) Experimental setup. The box ``Auto-Correlation'' is used here for measuring the marginal signal field using an HBT interferometer. (The setup ``Cross-Correlation'' is explored in Appendix~\ref{Sig-Idl-g2}.) 
(b) Joint spectral intensity of the BFC for CW pumping with an on-chip power of $\sim$10~mW. (c) Conceptual illustration of the proposed scheme. The box around the central signal frequency bin after modulation denotes a new, temporally magnified comb with an effective 1~GHz FSR.}
\label{Fig_1}
\end{figure*}

Figure~\ref{Fig_1}(a) illustrates the experimental setup. We use a Si$_3$N$_4$ microring, fabricated using the photonic Damascene reflow process~\cite{liu2020photonic,liu2021high}, to generate BFCs via SFWM. The resonator has a 2~$\upmu$m $\times$ 0.95~$\upmu$m cross-section and 561~$\upmu$m radius, corresponding to an FSR of $40.5$~GHz.   The gap between the ring and the bus waveguide is $0.3~\upmu$m, resulting in overcoupling with intrinsic and loaded $Q$-factors of $\sim$10$^7$ and $\sim$10$^6$, respectively.
An amplified CW laser operating at one resonance (1550.9~nm) pumps the MRR at power levels below the parametric threshold. The joint spectral intensity (JSI) measured under $\sim$10~mW of on-chip CW pump power is shown in Fig.~\ref{Fig_1}(b). The first two resonances are blocked by pump filters, leaving bins $\{3,4,...,51\}$ for the signal and $\{-51,-50,...,-3\}$ for the idler available for testing. The estimated on-chip pair generation rate varies between $\sim$0.5$\times$10$^6$ and $\sim$2.2$\times$10$^6$ s$^{-1}$ per frequency-bin par and the coincidence-to-accidental ratio (CAR)---defined here as the ratio of the average diagonal elements to the average of the off-diagonal elements---is $\sim$27.

In the case of pure CW pumping, the laser linewidth is much smaller than that of the MRR resonance, resulting in BFCs possessing time-energy-entanglement within any signal-idler bin pair. To examine states in which this intra-bin entanglement is removed, we also consider pump pulses with bandwidth exceeding the resonance linewidth, so that the effective number of Schmidt modes per bin pair approaches one~\cite{kues2017chip, vernon2017truly,vaidya2020broadband}. To generate these pump pulses, we carve the CW input with an an electro-optic intensity modulator (EOIM) biased at null transmission and driven with rectangular RF pulses. Dense wavelength-division multiplexing filters [not shown in Fig.~\ref{Fig_1}(a)] are placed before the ring to block amplified spontaneous emission and after it to suppress the residual pump. We then use a programmable pulse shaper~\cite{weiner2000femtosecond, Weiner2011} to select the BFC bins under test and potentially apply spectral phases, followed by an electro-optic phase modulator (EOPM) for sideband generation. A second pulse shaper functions as a wavelength selective switch (WSS) and routes the frequency bins to a 50:50 beamsplitter for HBT interferometry, where the two outputs of the beamsplitter are connected to SNSPDs. 

In light of the SFWM process, we expect to observe thermal statistics for the unheralded signal field~\cite{tapster1998photon, ou1999photon, mauerer2009colors, blauensteiner2009photon, christ2011probing, ou2017quantum, luo2015direct, luo2017temporal, kues2017chip,vaidya2020broadband}. 
We focus on auto-correlation measurements using an HBT interferometer, due to the wide interest therein for evaluating the spectral purity of entangled-photon sources~\cite{christ2011probing, kues2017chip, khodadad2021spectral, vaidya2020broadband}. However, our approach applies equally well to other time-resolved BFC measurements, such as the signal-idler cross-correlation shown as an inset; experimental examples of this configuration are provided in Appendix~\ref{Sig-Idl-g2}.

The narrow linewidth ($\sim$200~MHz) of the microring resonances allows for direct measurement of the temporal correlation functions for a single comb line. However, owing to the combined coincidence jitter of $110$~ps for our two SNSPDs, temporal features of multiline correlation functions 
are averaged out. To perform time-resolved measurements, 
we drive the EOPM with a sinusoid of RF frequency $39.5$~GHz, 
effectively generating a new set of BFC lines with an FSR of $1$~GHz, as shown in Fig.~\ref{Fig_1}(c) for the case of three frequency bins. 
We choose an RF modulation index of 1.43 rad on the EOPM in order to weight the contributions from the original frequency bins equally in the new BFC without any amplitude filtering. We present all the results below without accidental subtraction.

\section{Second-order coherence of the marginal signal field}\label{Sig-Sig g2}
The quantum state $\ket{\Psi}$ of our SFWM process can be written as~\cite{ou2017quantum}
\begin{multline}\label{Eq1}
    \ket{\Psi} \propto \int {\rm d}\omega_s {\rm d}\omega_i~\psi(\omega_s,\omega_i)\opdag{a}_s(\omega_s)\opdag{a}_i(\omega_i)\ket{0}\\ + ~\frac{1}{2} \int {\rm d}\omega_s {\rm d}\omega_i {\rm d}\omega_{s}' {\rm d}\omega_{i}'~\psi(\omega_s,\omega_i)\psi(\omega_{s}',\omega_{i}')
    \\ \times\opdag{a}_s(\omega_s)\opdag{a}_i(\omega_i)\opdag{a}_s(\omega_{s}')\opdag{a}_i(\omega_{i}')\ket{0}.
\end{multline}
Here $\psi(\omega_s,\omega_i)$ represents the joint-spectral amplitude (JSA) of a single biphoton, 
$\opdag{a}_{s(i)}(\omega_{s(i)})$ denotes the creation operator for a signal (idler) photon at frequency $\omega_{s(i)}$, and $\ket{0}$ is the vacuum state. The first term corresponds to the generation of a single signal-idler photon pair, whereas the second term corresponds to the simultaneous generation of two photon pairs. Higher-order terms are omitted as their contributions to the second-order temporal correlation functions are negligible compared to these first two. Because the theoretical model for SFWM results in the pure state of Eq.~\eqref{Eq1}, our analysis below specializes to pure states for theoretical calculations. However, the Vernier electro-optic approach proposed here applies equally well to mixed states, and indeed can provide insight into the phase coherence---and by implication, mixedness---of the joint signal-idler state via cross-correlation measurements (see Appendix~\ref{Sig-Idl-g2}).

The auto-correlation function of the marginal signal field $g^{(2)}_{ss}(t,t+\tau)$ for biphoton state $\ket{\Psi}$ can then be expressed as~\cite{ou2017quantum,mandel1995optical}:
\begin{multline}\label{Eq2}
    g^{(2)}_{ss}(t,t+\tau) = 
    \\ \frac{\bra{\Psi} \op{E}_{s}^{(-)}(t) \op{E}_{s}^{(-)}(t+\tau) \op{E}_{s}^{(+)}(t+\tau) \op{E}_{s}^{(+)}(t) \ket{\Psi}}
    {\bra{\Psi} \op{E}_{s}^{(-)}(t) \op{E}_{s}^{(+)}(t) \ket{\Psi} \bra{\Psi} \op{E}_{s}^{(-)}(t+\tau) \op{E}_{s}^{(+)}(t+\tau) \ket{\Psi}},
\end{multline}
where $\op{E}_{s}^{(+)}(t)= \frac{1}{\sqrt{2\pi}}\int {\rm d}\omega\op{a}_s(\omega)e^{-\imag\omega t}$ is the positive-frequency electric field operator for the signal field and $\op{E}_s^{(-)}(t)$ is its Hermitian conjugate; $t$ and $\tau$ are time variables, with $\tau$ representing the difference in arrival time of the two photons. The numerator corresponds to the probability density of measuring one signal photon at time $t$ and another at time $t+\tau$, so that only the second term in Eq.~\eqref{Eq1} contributes to to the numerator; by contrast, the factors in the denominator describe single-photon events and are hence dominated by the first term in Eq.~\eqref{Eq1}. 


\subsection{CW Pumping}

\begin{figure*}[!htb]
  \centering
  \includegraphics[width=\textwidth]{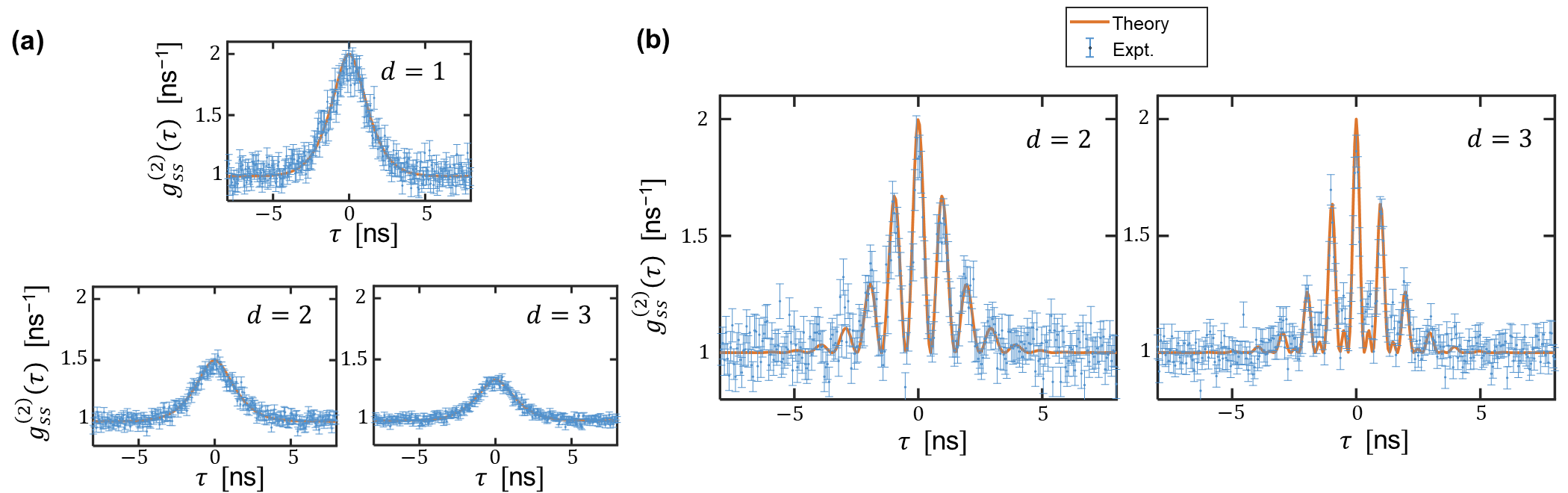}
\caption{Time-resolved auto-correlation measurements of the marginal signal field with $d$ bins under CW pumping. (a)~Without RF modulation. (b)~With RF modulation and spectral filtering.
The coincidence histograms used to compute $g^{(2)}_{ss} (\tau)$ in (a) and (b) are acquired over $T_\text{acq}=15$~min. and $T_\text{acq}=120$~min., respectively. The histogram bin width is $T_\text{bin}=64$~ps. Error bars assume Poissonian counting statistics.}
\label{Fig_2}
\end{figure*}

When seeded by a monochromatic CW pump, the JSA $\psi(\omega_s,\omega_i)$ in Eq.~\eqref{Eq1} factorizes into $\psi(\omega_s,\omega_i)=\varphi(\omega_s)\delta(\omega_s+\omega_i-2\omega_0)$, where $\omega_0$ is the pump frequency. If we consider a BFC, generated by a CW pump resonant with a cavity mode, of dimension $d$ possessing identical frequency lineshapes with a Lorentzian full-width at half-maximum (FWHM) $\gamma$ and FSR $\Delta\omega$, $\varphi(\omega_s)$ is given by \cite{chen2011frequency}:
\begin{equation}\label{Eq3}
    \varphi(\omega_s) \propto \sum_{k=k_0}^{k_0+d-1} \frac{\alpha_k}{(\frac{\gamma}{2})^2 + (\omega_s-\omega_k)^2 },
\end{equation}
where $k_0$ is a positive integer and $\omega_k = \omega_0 + k\Delta\omega$ defines the frequency grid. Thus, each term $k$ corresponds to a signal-idler pair shifted by $\pm k\Delta\omega$ from the pump, with complex probability amplitude $\alpha_k$. In this case,  the JSA of an individual signal-idler resonance pair is not spectrally factorable, i.e., it possesses intra-bin time-energy entanglement in addition to the frequency-bin entanglement present across multiple signal-idler bins-pairs. 

Equations~(\ref{Eq1}--\ref{Eq3}) together lead to the auto-correlation function 
\begin{equation}
\label{Eq4}
    g_{ss}^{(2)}(t,t+\tau) =1 + e^{-\gamma|\tau|}\Bigg(1 + \frac{\gamma|\tau|}{2}\Bigg)^{2}\Bigg| \frac{1}{d} \sum_{k=1}^{d} e^{\mathrm{i}k\Delta \omega\tau} \Bigg|^2, 
\end{equation}
under the assumption of equiprobable frequency bins ($|\alpha_k|^2=|\alpha_{k'}|^2\;\forall\;k,k'$) and well-separated resonances $\gamma \ll \Delta\omega$. A similar expression has been derived for SPDC~\cite{luo2015direct, luo2017temporal}. We note that $g^{(2)}_{ss}(t,t+\tau)\equiv g^{(2)}_{ss}(\tau)$ is independent of $t$, as expected for a stationary process. In addition, $g^{(2)}_{ss}(\tau)$ is also independent of the phase of the complex amplitudes $\alpha_k$. The lineshape parameter $\gamma$ defines the width of the exponentially decaying envelope, and $\Delta\omega$ dictates the multimode interference pattern within the envelope. The peak value of $g^{(2)}_{ss}(0)=2$ is indicative of photon bunching from the thermal mode statistics of the unheralded signal field~\cite{ou2017quantum,mandel1995optical}.

For the CW experiments, we consider a state with $k_0=12$ and $d\in\{1,2,3\}$, corresponding to signal bins 12, 13, and 14 in Fig.~\ref{Fig_1}(b). Bypassing the EOIM in Fig.~\ref{Fig_1}(b), we pump the MRR directly with a CW bus waveguide power of $\sim$14~mW, well below the $\sim$80~mW parametric threshold power for classical comb generation. 
We compute the normalized second-order auto-correlation, from the raw counts recorded in our histogram integrated over an acquisition time $T_\text{acq}$, as
\begin{equation}\label{CW-g2-expt}
g^{(2)}_{ss}(\tau=rT_\text{bin}) = \frac{T_\text{acq}}{T_\text{bin}}\frac{N_{ab}(r)}{N_a N_b},    
\end{equation} where $N_{ab}(r)$ is the total number of coincidences in the $r^\text{th}$ histogram bin with time-bin width $T_\text{bin}$ (64~ps in our experiment) at a relative delay $\tau = rT_\text{bin}$; $N_a$ and $N_b$ are the total singles counts registered in the two detectors over $T_\text{acq}$. 

First, we do not apply any RF modulation on the EOPM and pass the signal bins of interest from the WSS to the HBT interferometer. These results are shown in Fig.~\ref{Fig_2}(a). The SNSPD jitter averages out the multimode interference fringes occurring in the timescale of $2\pi/\Delta\omega = 24.7$~ps, leading to measured peak values of $g_{ss}^{(2)}(0)\in\{1.51\pm0.05,1.33\pm0.03\}$ for $d\in\{2,3\}$, in good agreement with the theoretical prediction of 
$g_{ss}^{(2)}(0) = 1 + \frac{1}{d}$ under temporal averaging. Thus, while the $g_{ss}^{(2)}(0)=2$ bunching associated with thermal statistics is clearly observed for $d=1$, temporal averaging leads to reduced peak values for higher dimensions, from which one could falsely conclude coherent statistics [$g^{(2)}(0)\rightarrow1$]. 

Next, we employ our RF phase modulation scheme to reveal the previously hidden fine features in $g^{(2)}_{ss}(\tau)$. We drive the EOPM with a 39.5~GHz sinusoid and pass the signal photons around the central frequency bin [highlighted by the signal box in Fig.~\ref{Fig_1}(c)] to the HBT interferometer. The experimental results for $d=2$ and $d=3$ appear in Fig.~\ref{Fig_2}(b), 
showing excellent agreement with the plotted theoretical prediction, namely, Eq.~\eqref{Eq4} with $\gamma$ and $\Delta\omega$ fit to minimize squared error. From the fits,   $\Delta\omega/2\pi$ ranges between $1.006$~and~$1.007$~GHz (effective FSR of the BFC after phase modulation) and $\gamma/2\pi$ ranges between $236$~and~$264$~MHz. Note also that the individual beat features narrow from FWHMs of 480~ps for $d=2$ to 255~ps for $d=3$, consistent with the increase in effective BFC bandwidth from 2 to 3~GHz. Importantly, our scheme allows for the direct observation of thermal mode statistics [$g_{ss}^{(2)}(0)=2]$ for the superposition of multiple frequency bins---a feature previously lost in the averaged-out results of Fig.~\ref{Fig_2}(a).

References~\cite{guo2017parametric, samara2021entanglement, yamazaki2022massive, zhang2021chip} have all measured time-resolved $g^{(2)}_{ss}(\tau)$ for a single resonance in SPDC or SFWM. However, when multiple comb lines have been considered, previous works have reported only washed-out interferograms~\cite{luo2015direct,luo2017temporal, yamazaki2022massive,zhang2021chip}. Our scheme effectively improves upon these limitations, enhancing the temporal resolution for time-resolved measurement of multiple comb lines and enabling direct confirmation of thermal statistics [$g_{ss}^{(2)}(0)=2$].

\subsection{Pulsed Pumping}

\begin{figure*}[!htb]
  \centering
  \includegraphics[width=\textwidth]{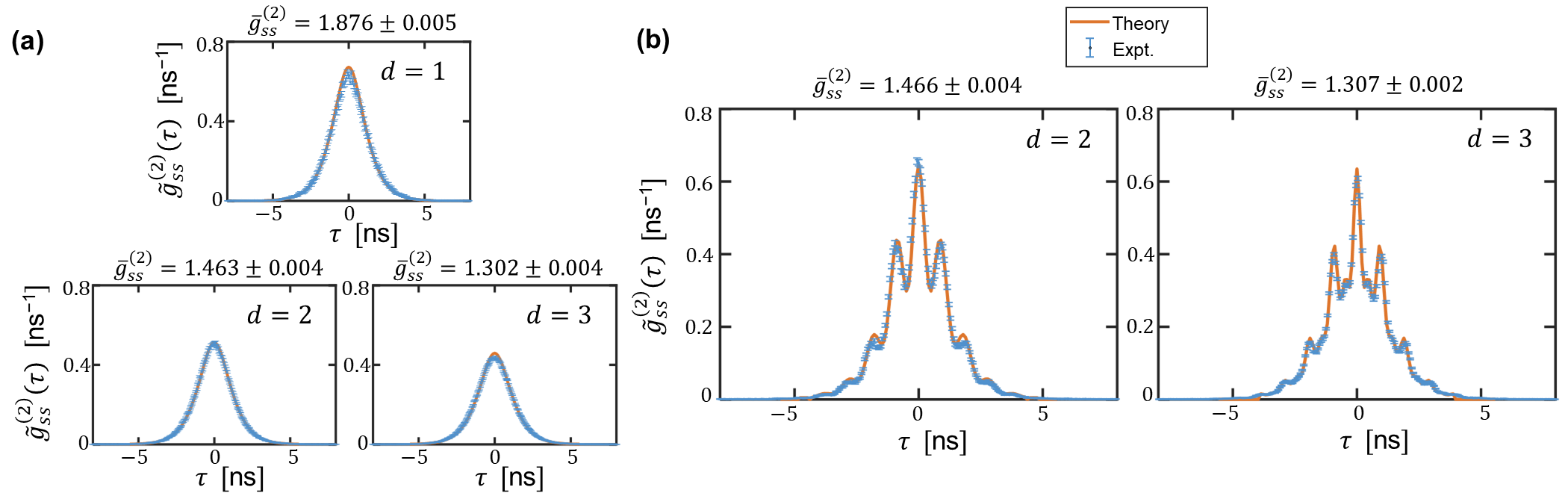}
\caption{Time-resolved auto-correlation density measurements [cf. Eq.~\ref{g2_density_def}] of the marginal signal field with $d$ bins under pulsed pumping. (a) Without RF modulation. (b) With RF modulation and spectral filtering. 
The coincidence histograms used to compute $\Tilde{g}^{(2)}_{ss} (\tau)$ in (a) and (b) are acquired over $3$~min. and $15$~min., respectively. The histogram bin width is $T_\text{bin}=64$ ps. The time-integrated $\overline{g}^{(2)}_{ss}$ values representing the area under the experimental correlation density are included in the plots. Error bars assume Poissonian statistics.}
\label{Fig_3}
\end{figure*}

We now explore the scenario where SFWM is seeded by a pulsed pump whose spectrum fills an entire microring resonance. Pulsed pumping is often pursued for producing spectrally pure single photons by erasing the intra-bin time-energy entanglement~\cite{eckstein2011highly, kaneda2016heralded}.  
If the coincidence window encompasses the pulse duration, then an \emph{integrated} auto-correlation function of the unheralded signal photons from Eq.~\eqref{Eq2} can be defined for isolated pump pulses as
\begin{equation}\label{Int_g2}
    \overline{g}^{(2)}_{ss} = \frac{\iint {\rm d}t~{\rm d}\tau ~G^{(2)}(t, t+\tau)}{\left[\int {\rm d}t~G^{(1)}(t)\right]^2},
\end{equation}
where
the correlation functions are defined as
\begin{multline}
G^{(2)}(t, t+\tau) \\  = \bra{\Psi} \op{E}_{s}^{(-)}(t) \op{E}_{s}^{(-)}(t+\tau) \op{E}_{s}^{(+)}(t+\tau) \op{E}_{s}^{(+)}(t) \ket{\Psi}
\end{multline}
and
\begin{equation}\label{Ps}
    G^{(1)}(t) = \bra{\Psi} \op{E}_{s}^{(-)}(t) \op{E}_{s}^{(+)}(t) \ket{\Psi}.
\end{equation}
In words, the integrals appearing in the numerator and denominator of Eq.~\eqref{Int_g2} represent the two-photon and single-photon probability per pulse, respectively. 
The time-integrated $\overline{g}^{(2)}_{ss}$ is often used to infer the Schmidt number $K$ of the biphoton JSA per frequency bin, and consequently its factorability, via the relation $\overline{g}^{(2)}_{ss} = 1+ \frac{1}{K}$
~\cite{christ2011probing,ou2017quantum}. For example, a measurement of $\overline{g}^{(2)}_{ss} = 2$ implies that the JSA comprises a single Schmidt mode and can be expressed as a fully-separable product state, $\psi(\omega_s, \omega_i) = \varphi_s(\omega_s)\varphi_i(\omega_i)$. However, measuring $d$ frequency bins, with $K$ Schmidt modes each, simultaneously results in a scaling of the form $\overline{g}^{(2)}_{ss} = 1+ \frac{1}{dK}$.
To illuminate 
the temporal and spectral features that are lost in $\overline{g}^{(2)}_{ss}$, we introduce the second-order auto-correlation \emph{density} $\Tilde{g}^{(2)}_{ss}(\tau)$ to probe the marginal signal field under pulsed pumping, defined as:
\begin{equation}\label{g2_density_def}
    \Tilde{g}^{(2)}_{ss}(\tau) = \frac{\int {\rm d}t~G^{(2)}(t, t+\tau)}{\left[\int {\rm d}t~G^{(1)}(t)\right]^2}
 \end{equation}
Unlike the fully time-integrated $\overline{g}^{(2)}_{ss}$ in Eq.~\eqref{Int_g2}, here we integrate $G^{(2)}(t, t+\tau)$ only over the $t$ variable, 
thus resulting in a correlation density with units of inverse time. The area under $\Tilde{g}^{(2)}_{ss}(\tau)$ returns $\overline{g}^{(2)}_{ss}$.

Using Eq.~\eqref{g2_density_def} and the state in Eq.~\eqref{Eq1}, the second-order auto-correlation density takes the following form: 
\begin{widetext}
\begin{equation}\label{g2_density_fin}
\begin{aligned}
    \Tilde{g}^{(2)}_{ss}(\tau) = \frac{ \iiiiint {\rm d}\omega_1 {\rm d}\omega_2 {\rm d}\omega'_1 {\rm d}\omega'_2 {\rm d}\omega_3~ \psi^{*}(\omega_1,\omega_2 ) \psi^{*}(\omega'_1,\omega'_2 )\psi(\omega_1+\omega'_1-\omega_3,\omega_2) \psi(\omega_3,\omega'_2) e^{-\imag\omega_3\tau}(e^{\imag\omega'_1\tau} + e^{\imag\omega_1\tau})} {2\pi \left(\iint {\rm d}\omega_1 {\rm d}\omega_2\left| \psi(\omega_1,\omega_2) \right|^2\right)^2}.
\end{aligned}
\end{equation}
\end{widetext}




When a pulse with spectral amplitude $\alpha_p(\omega)$ is used to pump the Lorentzian resonance $l_0(\omega)$ of the MRR, centered at $\omega_0$, the resultant JSA $\psi(\omega_s, \omega_i)$ 
due to SFWM can be expressed as~\cite{vernon2017truly, christensen2018engineering}
\begin{equation}\label{JSA_SWFM}
\psi(\omega_s, \omega_i) \propto \sum_{k=k_0}^{k_0+d-1} F_p (\omega_s + \omega_i) l_{k}(\omega_s) l_{-k}(\omega_i),
\end{equation}
where
\begin{equation}\label{Fp_SWFM}
F_p (\omega)  = \int {\rm d}\omega_p~\alpha_p(\omega_p) \alpha_p(\omega- \omega_p) l_0(\omega_p)l_0(\omega-\omega_p)
\end{equation}
and $l_n(\omega) \propto [\gamma/2 - \imag(\omega- \omega_n)]^{-1}$ defines a Lorentzian lineshape function centered at frequency $\omega_n = \omega_0 + n\Delta\omega$. 
Here we assume the bandwidth of the pump pulse is much less than the FSR ($\Delta\omega$) of the MRR.

We now experimentally measure $\Tilde{g}^{(2)}_{ss}(\tau)$
on pumping the MRR with pulses 
created by driving the EOIM in Fig.~\ref{Fig_1}(a) with 500~ps-wide RF waveforms repeating every 25~ns; the average on-chip optical power is 30~mW.
We experimentally compute $\Tilde{g}_{ss}^{(2)}(\tau)$ from our coincidence histogram with bin size $T_\text{bin}$ as
\begin{equation}\begin{aligned}\label{g2-density-expt}
   \Tilde{g}_{ss}^{(2)}(\tau = rT_\text{bin}) = \frac{T_\text{acq}}{T_\text{bin}T_\text{rep}}\frac{N_{ab}(r)}{N_a N_b},
\end{aligned}
\end{equation}
where $T_\text{rep}$ denotes the repetition period of the pulses. Compared to Eq.~\eqref{CW-g2-expt}, the extra factor $1/T_\text{rep}$ leads to a dimensioned quantity whose time-integral is dimensionless.



We perform measurements for up to $d=3$ signal bins and $k_0=21$ for another cluster of resonances [21, 22, and 23 in Fig.~\ref{Fig_1}(b)], with the results in Fig.~\ref{Fig_3}. 
The smaller error bars compared to those in Fig.~\ref{Fig_2} stem from a larger number of coincidence counts: for SFWM, the coincidence rate for two signal photons scales quarticly with the peak pump power, which is significantly higher in these pulsed tests.
In the absence of phase modulation [Fig.~\ref{Fig_3}(a)], the curves found correspond to the integrated values $\overline{g}^{(2)}_{ss} \in\{1.88,1.46,1.3\}$ for $d\in\{1,2,3\}$. The detector jitter averages out the interference fringes occurring due to cavity modes spaced at the 40.5~GHz FSR for $d=2$ and $3$, resulting  in a reduced peak in Fig.~\ref{Fig_3}(a) 
that scales in proportion to $\overline{g}^{(2)}_{ss}$, which are in close agreement with the expected trend $\overline{g}^{(2)}_{ss} = 1 + \frac{1}{dK}$ taking $K=1.14$. 

The theoretical curves are obtained through numerical integration of Eq.~\eqref{g2_density_fin} using $\gamma/2\pi = 200$~MHz (found in previous work~\cite{lu2022bayesian}) and assuming a Gaussian optical pump spectrum with a 1.1~GHz FWHM (obtained from temporal characterization of the input pulses and assumed transform-limited),
from which we compute $\overline{g}^{(2)}_{ss}=1.9$ for $d=1$---close to the measured value of 1.88. Importantly, apart from the FSR, resonance, and pump bandwidths, no other inputs are provided to the theoretical model; in particular, no vertical scaling is applied to the theory curves in Fig.~\ref{Fig_3}(a), so that the absolute agreement provides meaningful verification.
Both $\overline{g}^{(2)}_{ss}$ and $\Tilde{g}^{(2)}_{ss}(\tau)$ in Fig.~\ref{Fig_3}(a) are incapable of capturing the fine temporal features and information regarding the thermal behavior of multiple frequency bins is hence lost. 
As an aside, we note that the maximum attainable integrated auto-correlation for a Gaussian-pulse-pumped microring BFC is $\overline{g}^{(2)}_{ss}=1.92$~\cite{vernon2017truly}---only slightly higher than the 1.9 predicted (and 1.88 measured) in our experiment,  indicating operation close to the optimal regime for separability. (Integrated $\overline{g}^{(2)}_{ss}>1.92$ for a microring source is possible only with 
special device engineering~\cite{vernon2017truly,liu2020high} or complex pump pulse tailoring to achieve a flat $F_p(\omega)$ [Eq.~\eqref{Fp_SWFM}]~\cite{christensen2018engineering,burridge2020high}.)

We now employ 
phase modulation at 39.5~GHz to recover the lost features in $\Tilde{g}_{ss}^{(2)}(\tau)$. The results are presented in Fig.~\ref{Fig_3}(b) and are again in close agreement with theory, capturing the temporal features that reveal the number of frequency bins $d$ in the BFC state. Furthermore, the peak value of the correlation density $\Tilde{g}_{ss}^{(2)}(\tau=0)$ is no longer sensitive to the number of bins, similar to the CW pumping scenario. 
Our approach thus not only provides insight into the Schmidt number of the source like fully integrated HBT measurements \cite{christ2011probing,kues2017chip,vaidya2020broadband} but also reveals fine temporal features resulting from the number and shape of the contributing frequency bins.

Interestingly, the presence of central $\tau=0$ spikes and lower pedestals in Fig.~\ref{Fig_3}(b) directly matches known effects in classical ultrafast optics as well. In auto-correlation measurements of a pulsed source, the presence of a short peak protruding above a longer finite-duration base is the key signature of a noisy field with bandwidth appreciably wider than the inverse pulse duration~\cite{pike1970basis, Ippen1984, Weiner2009}. 
Our findings with photon counting bear a remarkable resemblance to this form; the presence of \emph{multiple} temporal peaks---rather than a single one at $\tau=0$---derives from the discrete nature of the comblike thermal spectrum in our case.

\begin{figure*}[!tb]
  \centering
  \includegraphics[trim=0 10 190 0,clip,width=0.7\textwidth]{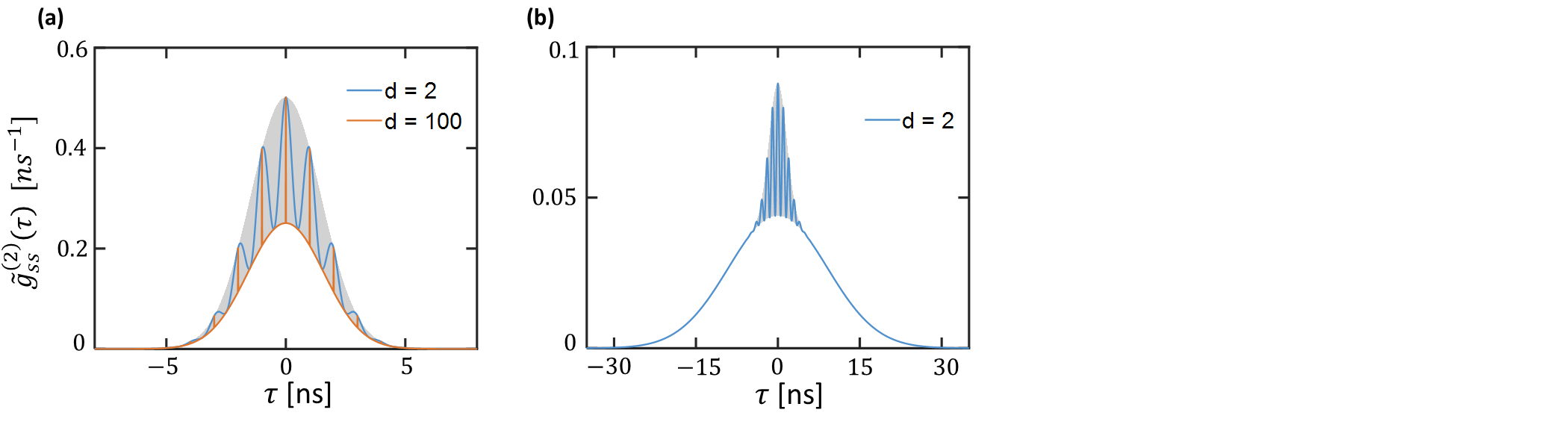}
\caption{Second-order correlation density $\Tilde{g}^{(2)}_{ss} (\tau)$  of the marginal signal field for BFCs given in Eq.\eqref{Eq18}. $\Delta \omega/2\pi = 1$~GHz is assumed. (a) $\sigma_p/2\pi = \sigma_r/2\pi = 0.2$~GHz (b) $\sigma_p/2\pi = 0.025$~GHz and $\sigma_r/2\pi = 0.2$~GHz. The details regarding the shaded region are given in the text.} 
\label{Fig_6}
\end{figure*}

The connection to ultrafast optics can be elucidated by exploring Gaussian spectra for analytical evaluation. Consider a JSA written as
\begin{multline}
\label{Eq18}
    \psi(\omega_s,\omega_i) \propto \sum_{k=k_0}^{k_0+d-1} e^{-(\omega_s + \omega_i - 2\omega_0)^2/\sigma_p^2} \\ \times e^{-[(\omega_s - \omega_k)^2+(\omega_i - \omega_{-k})^2]/\sigma_r^2},
\end{multline}
where $\sigma_p$ and $\sigma_r$ are related to the pump and resonance linewidths, respectively. The resonance frequencies $\omega_{\pm k}$ are as defined before. The state of this form is very close to that in Eq.~\eqref{JSA_SWFM}, but with Gaussian functions in place of $F_p$ and $l_n$. Plugging Eq.~\eqref{Eq18} into Eq.~\eqref{g2_density_fin} and assuming $\sigma_p,\sigma_r \ll \Delta\omega$ (the typical case for a BFC) leads to a convenient closed-form expression for $\Tilde{g}^{(2)}_{ss}(\tau)$:
\begin{widetext}
\begin{equation}\label{Eq19}
    \Tilde{g}^{(2)}_{ss}(\tau) = \frac{\sigma_r}{\sqrt{4\pi\left( 1 + \frac{\sigma_r^2}{\sigma_p^2} \right)}} e^{-\frac{\sigma_r^2\tau^2}{4\left( 1 + \frac{\sigma_r^2}{\sigma_p^2} \right)}} \left\{ 1    
     + e^{-\frac{\sigma_r^4\tau^2}{2\sigma_p^2\left( 1 + \frac{\sigma_r^2}{\sigma_p^2} \right)}} \frac{1}{d^2} \left| \sum_{k=k_0}^{k_0+d-1} e^{ik\Delta\omega \tau} \right|^2 \right\},
\end{equation}
\end{widetext}
which follows the form $\Tilde{g}_{ss}^{(2)}(\tau) = \Gamma(\tau) [1 + |\lambda(\tau)|^2]$, where $\Gamma(\tau)$ is a finite-duration envelope and $|\lambda(\tau)|\leq 1$ describes the coherence spikes. Such an expression matches precisely that in simple classical noise analyses~\cite{pike1970basis}, but now derived directly from a biphoton JSA $\psi(\omega_s,\omega_i)$. The envelope $\Gamma(\tau)$ defines a lower pedestal and satisfies $\int {\rm d}\tau~\Gamma(\tau)=1$ for any parameter settings; $\lambda(0)=1$, creating the 2:1 contrast at $\tau=0$. For $d=1$, $\lambda(\tau)$ contains no oscillations and can be viewed as an upper envelope that bounds any interference fringes for $d>1$ cases.

In Fig.~\ref{Fig_6}(a) we plot Eq.~\eqref{Eq19} for $\sigma_p/2\pi = \sigma_r/2\pi = 0.2$~GHz and $\Delta \omega/2\pi = 1$~GHz (similar to the bandwidths experimentally tested in Fig.~\ref{Fig_3}), but now for two extreme cases of frequency-bin number: $d\in\{2,100\}$. 
The peaks are significantly narrower for $d=100$ compared to $d=2$, leading to a reduced area under the curve such that $\overline{g}^{(2)}_{ss} = 1 + \frac{1}{dK}$ holds, where $K=1.14$ is the Schmidt number for a single pair of lines.
As an example when $\sigma_r$ and $\sigma_p$ differ strongly, we also plot $\Tilde{g}_{ss}^{(2)}(\tau)$  for  $\sigma_p/2\pi = 0.025$~GHz, $\sigma_r/2\pi = 0.2$~GHz, and $d=2$ in Fig.~\ref{Fig_6}(b). The lower envelope now widens significantly compared to the upper one, a situation occurring in practice when a BFC is pumped by a pulse much longer than the inverse microring lineshape. We obtain a higher value of $K=5.88$ as the Schmidt number, as expected for a narrowband pump ($\sigma_p<\sigma_r$). 
Accordingly, this simple Gaussian model offers an intuitive picture into experimental findings with the auto-correlation density, illuminating interesting connections to short-pulse characterization in classical optics~\cite{pike1970basis, Ippen1984, Weiner2009}.

Finally, to wrap up this section, we note that a peak value of $g^{(2)}_{ss}(0)=2$ in the case of CW pumping is completely different from the time-integrated condition $\overline{g}^{(2)}_{ss}=2$ in pulsed pumping. The time-resolved result $g^{(2)}_{ss}(0)=2$ in CW pumping is an indicator of the thermal nature of the marginal signal and is not related to the Schmidt number $K$ of the original biphoton source, whereas $\overline{g}^{(2)}_{ss}=2$ implies $K=1$ for an SFWM or SPDC source. However, these two distinct quantities have been often confused in the previous works, where $g^{(2)}_{ss}(0)=2$ in CW pumping has been misinterpreted as implying $K=1$ ~\cite{zhang2021chip,samara2021entanglement,guo2017parametric}. 
We therefore emphasize the importance in distinguishing between these measurement scenarios, particularly in detector-jitter-impacted contexts.


\section{Conclusion}

We have presented an experimental scheme based on Vernier electro-optic modulation to directly resolve fine temporal features in BFC correlation functions. Our accompanying theoretical framework encompasses both CW and pulsed pumping scenarios and attains excellent agreement with experiment. While our concentration has been on the particular case of HBT auto-correlation measurements, the technique applies generally to a variety of time-resolved biphoton characterization methods; for example, Appendix~\ref{Sig-Idl-g2} offers additional results for signal-idler cross-correlation measurements performed in our setup.


In many ways, this Vernier electro-optic technique furnishes a bridge between (i)~the simple---but relatively high-jitter---approach of direct single-photon detection and (ii)~nonlinear optical approaches that offer femtosecond-scale resolution, but at the cost of comparatively low efficiency and high experimental complexity~\cite{Peer2005,boitier2009measuring, Lukens2013c,boitier2013two, Kuzucu2008b, MacLean2018, joshi2022picosecond}. Nevertheless, continued advances in single-photon detectors are rapidly closing the speed gap between SNSPDs and EOPMs. Indeed, few-picosecond jitters have been demonstrated at telecom wavelengths in state-of-the-art SNSPDs~\cite{Korzh2020}, so that direct observation of $\sim$40~GHz fringes seems feasible with existing technology. Accordingly, the regime in which Vernier modulation can enhance temporal resolution will depend on \emph{both} detector and EOPM properties; whenever the bandwidth of available EOPMs exceeds that of available single-photon detectors---as in the experimental examples here---our approach applies, adding one more powerful tool to the broad field of quantum-optical characterization. 

\section*{Acknowledgments} 
This research was performed in part at Oak Ridge National Laboratory, managed by UT-Battelle, LLC, for the U.S. Department of Energy under contract no. DE-AC05-00OR22725. Funding was provided by the U.S. Department of Energy (
ERKJ353); the National Science Foundation (1839191-ECCS, 2034019-ECCS); the Air Force Office of Scientific Research (
FA9550-19-1-0250); and the Swiss National Science Foundation (
176563). K.V.M. acknowledges support from the QISE-NET fellowship program of the National Science Foundation (DMR-1747426).
M.S.A. acknowledges support from the College of Engineering Research
Center at King Saud University. The Si$_3$N$_4$ samples were fabricated in the
EPFL Center of MicroNanoTechnology (CMi). We thank D.~E. Leaird for technical assistance.

Portions of this work were presented at CLEO 2021 (paper number JM3F.5) and CLEO 2022 (paper number FTu5A.2).


\appendix

\section{Second-order cross-correlation measurements} \label{Sig-Idl-g2}

\begin{figure*}[!htb]
  \centering
  \includegraphics[width=\textwidth]{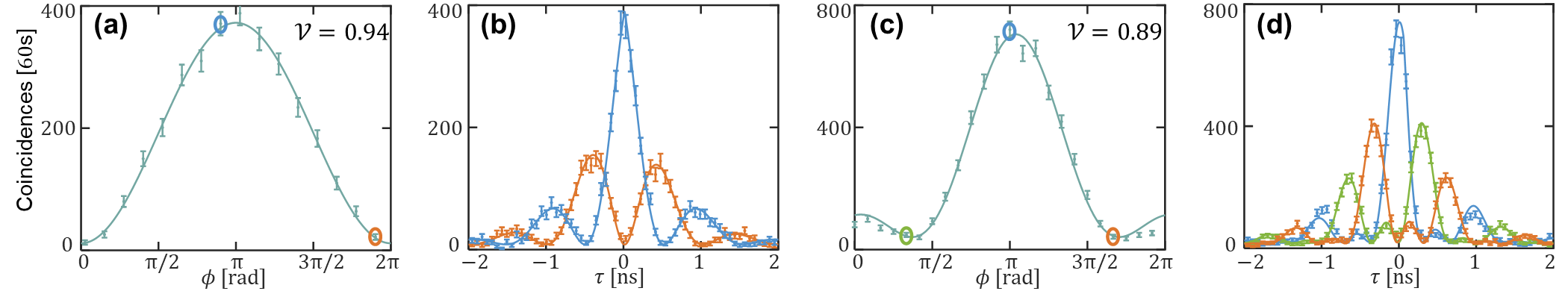}
\caption{ (a)~Measured signal-idler coincidences at zero delay ($\tau=0$) as spectral phase $\phi$ is swept for $d=2$ pairs of bins. (b) Time-resolved cross-correlation functions obtained via Vernier modulation at the phases circled in (a). (c,d) Same as (a,b) but now for $d=3$. The visibilities of the interference curves are given in (a) and (c). The coincidence histograms are integrated over $T_\text{acq}=60$~s with a bin width of $T_\text{bin}=64$~ps. Error bars assume Poissonian counting statistics.}
\label{Fig_4}
\end{figure*}

To complement the focus on time-resolved auto-correlation measurements in the main text, here we provide an example applying our technique for another measurement configuration: the second-order \emph{cross}-correlation [see inset in Fig.~\ref{Fig_1}(a)]. Defined as
\begin{multline}\label{Eq15}
    g^{(2)}_{si}(t,t+\tau) = 
    \\ \frac{\bra{\Psi} \op{E}_{s}^{(-)}(t) \op{E}_{i}^{(-)}(t+\tau) \op{E}_{i}^{(+)}(t+\tau) \op{E}_{s}^{(+)}(t) \ket{\Psi}}
    {\bra{\Psi} \op{E}_{s}^{(-)}(t) \op{E}_{s}^{(+)}(t) \ket{\Psi} \bra{\Psi} \op{E}_{i}^{(-)}(t+\tau) \op{E}_{i}^{(+)}(t+\tau) \ket{\Psi}},
\end{multline}
this cross-correlation can be used to probe the phase coherence present across the multiple frequency bins and, in turn, the entanglement in the BFC state. We experimentally measure the coincidence rate between the signal and idler photons, 
again applying Vernier phase modulation to rescale the FSR down to the few-GHz range required for direct observation with single-photon detectors.

Because cross-correlation measurements are sensitive to spectral phase, in these experiments we use the pulse shaper to compensate dispersion and apply additional linear phases on the signal-idler spectra of the form $\beta_s[k]=\frac{\phi}{2}(k-k_0)$ for the signal bins $\omega_k$ and $\beta_i[-k]=\frac{\phi}{2}(k-k_0)$ for the idler bins $\omega_{-k}$, where $k\in\{k_0,...,k_0+d-1\}$. This V-shaped phase pattern introduces a controllable delay $\tau_0=\phi/\Delta\omega$ between the signal and idler photons~\cite{Peer2005}. For an SFWM process seeded by a CW pump, the signal-idler coincidence rate after modulation is given by
\begin{equation}
\label{eq:Rd}
    R_{si}(\tau) \propto g^{(2)}_{si}(\tau) \propto e^{-\gamma|\tau|}\left| \sum_{k=k_0}^{k_0+d-1} (-1)^k e^{\imag(k-k_0)(\phi-\Delta\omega \tau)} \right|^2,
\end{equation}
where we assume equal weights for each bin (amplitude equalization with a pulse shaper being required when $d>3$); 
$\Delta\omega/2\pi$ refers to the effective frequency separation in the BFC after phase modulation (1~GHz in our experiments). 

For the experiments in this section, we used another MRR of slightly different cross-section ($1.9~\upmu$m$\times$ $0.95~\upmu$m) and 40.4~GHz FSR. 
The amplified CW laser at 1550.9~nm pumps the MRR with an on-chip bus waveguide power of $\sim$14~mW, below the parametric threshold of $\sim$143~mW; the estimated on-chip pair generation rate is between 0.9--1.9$\times$10$^{6}$ s$^{-1}$ per frequency-bin pair, with a CAR of 30. We worked with BFCs up to $d=3$ (resonances $k\in\{3,4,5\}$) and drove the EOPM at $39.4$~GHz. 

Figure~\ref{Fig_4}(a) plots the coincidences measured at zero delay $R_{si}(0)$ for a 60~s integration time as $\phi$ is scanned. Dispersion has been compensated using the method described in Ref.~\cite{lu2022bayesian}.
Full time-resolved cross-correlation measurements for the two circled phases appear in Fig.~\ref{Fig_4}(b).
As in the CW auto-correlation tests with Vernier modulation [Fig.~\ref{Fig_2}(b)], clear fringes appear with a 1~ns period, but now with a contrast no longer limited to 2:1.  Adding one more combline to reach $d=3$, we obtain the zero-delay fringe and example time-resolved cross-correlation functions as plotted in Fig.~\ref{Fig_4}(c) and (d), respectively. In all cases, excellent agreement is obtained with the theoretical fits from Eq.~\eqref{eq:Rd}, from which  
we extract $\gamma/2\pi$ ranging between 307--363~MHz and $\Delta\omega/2\pi$ ranging between 0.96--0.99~GHz. Once again, the individual peaks narrow as $d$ increases. Moreover, for a given $d$, the total area under each $R_d(\tau)$ example curve [Fig.~\ref{Fig_4}(b,d)] remains constant, as the various cases differ only in spectral phase (and not amplitude).

The observation of fine features in the temporal correlation functions and its good agreement with the theoretical model serves as an indicator of entanglement present across the frequency bins. The visibility of the interference curves in Fig.~\ref{Fig_4}(b) and (d) are 0.94 and 0.89, respectively, which---under the assumption of isotropic noise---exceed the thresholds for Bell inequality violation (0.71 and 0.77 for $d=2$ and $d=3$, respectively)~\cite{thew2004bell, imany2018characterization}. Importantly, our proposed scheme helps probe the phase coherence and hence the entanglement in the BFC through the effectively improved temporal resolution.

\bibliography{References}

\end{document}